\documentclass[12pt]{iopart}
\usepackage{iopams} % ipv the 4 ams packages
\usepackage{cite}
\usepackage{hyperref}
\usepackage{graphicx}
\newcommand{\wigner}[6]{\left(\begin{array}{ccc} #1 & #2 & #3 \\ #4 & #5 & #6 \end{array}\right)} 
\begin{document}
\title{The quadrupole collective model from a Cartan-Weyl perspective.}
\author{S.~De~Baerdemacker$^{1,2}$, K.~Heyde${}^1$ and V.~Hellemans${}^1$}
\address{$^1$ Universiteit Gent, Vakgroep Subatomaire en Stralingsfysica, Proeftuinstraat 86, B-9000 Gent, Belgium\\
$^2$ Department of Physics, University of Toronto, Toronto, Ontario M5S 1A7, Canada}
\ead{stijn.debaerdemacker@ugent.be}
\begin{abstract}
The matrix elements of the quadrupole variables and canonic conjugate momenta emerging from collective nuclear models are calculated within an $SU(1,1)\times O(5)$ basis.  Using a harmonic oscillator implementation of the $SU(1,1)$ degree of freedom, one can show that the matrix elements of the quadrupole phonon creation and annihilation operators can be calculated in a pure algebraic way, making use of an intermediate state method.
\end{abstract}
\submitto{\JPA}
\maketitle

\section{Introduction}
%=====================
%
Collective modes of motion have proven to be very important in the study of low-energy spectra of medium and heavy-mass atomic nuclei, in regions with many valence protons and neutrons outside of the closed-shell regions \cite{bohr:98}.  In these regions, the early nuclear shell-model \cite{haxel:49,mayer:49,mayer:50} was not able to correctly describe the observed large quadrupole deformations, starting from a single-particle assumption.  It was suggested that all nucleons in the atomic nucleus cooperate in a collective dynamical way, thereby producing the deformation of the nucleus, due to polarisation effects in the nuclear medium \cite{rainwater:50}.  These ideas subsequently led to the development of the Bohr-Mottelson collective model, in which a single nucleon is coupled to the soft quadrupole deformed surface of the atomic nucleus \cite{bohr:52,bohr:53}.  The dynamics of the surface is governed by the Bohr Hamiltonian, describing vibrational and rotational-like excitation modes, depending on the potential energy term in the Hamiltonian (\ref{hamiltonian:hamiltonian}).\newline
One could start from an analytically solvable potential for the description of the collective properties of atomic nuclei.  This topic recently gained considerable attention thanks to its application in critical point descriptions of quantum shape phase transitions \cite{iachello:00,iachello:01,iachello:03}.  However, the class of analytically solvable potentials is restricted to a number of schematic and benchmarking cases.  For a recent overview on solvable potentials, we would like to refer the reader to \cite{fortunato:05a} and references therein.  In order to describe more general structures of collectivity, a numerical treatment of more general types of potentials is needed.  For this purpose, one requires a suitable basis to diagonalise the Hamiltonian.  Pioneering work was carried out by B\`es \cite{bes:58}, who constructed all wavefunctions of a $\gamma$-independent system up to angular momentum $L=6$ by means of a coupled differential equation technique.   Later on, a number of strategies were proposed to construct convenient bases, profoundly relying on the algebraic $SU(1,1)\times O(5)$ structure, which is inherent in the collective quadrupole model.  However, the angular momentum symmetry, which is observed in experimental nuclear structure data, is not contained in the natural Cartan-Weyl \cite{cartan:1894,wybourne:74,iachello:06} reduction of $O(5)$.  Hence, the $O(3)\supset O(2)$ symmetry has to be imposed explicitly on the basis wavefunctions.  One can either start from basic building blocks with good tensorial properties \cite{corrigan:76,chacon:76,chacon:77}, or from a projective coherent state procedure \cite{gheorghe:78} with good orthogonality properties \cite{szpikowski:80,gozdz:80}.  Also the vector coherent state formalism \cite{rowe:94a,rowe:94b,rowe:95,turner:06} and the more recently proposed algebraic tractable model \cite{rowe:04,rowe:05a} provide a basis with good angular momentum symmetry by construction, the latter already being used by Caprio in his examination of the $\beta-\gamma$ decoupling within the $X(5)$ critical point description\cite{caprio:05a}.\newline
Nevertheless, although the Cartan-Weyl reduction of $O(5)$ is not compatible with the $O(3)$ symmetry, it offers a convenient basis to calculate all necessary matrix elements of the collective variables in an algebraic straightforward way \cite{debaerdemacker:07a}. However, in the mentioned work, the matrix elements of the collective variables are calculated only in the $O(5)$ basis, leaving a freedom of choice for a suitable $SU(1,1)$ basis by considering the 5-dimensional radial variable $\beta$ as a constant of motion.  In the present work, we will eliminate this choice and demonstrate that the radial degree of freedom can be included in a similar straightforward way, using the intermediate state method.
\section{The collective model and its algebraic structure}
%==============================================================
%
The collective model starts from the assumption that the atomic nucleus exhibits a well-defined surface.  This surface is subject to deformations which can be described by means of a multipole expansion
\begin{equation}\label{ellipsoid:equation}
R(\theta,\phi)=R_0\Big(1+\sum_{\lambda\mu}\alpha^\ast_{\lambda\mu}Y_{\lambda\mu}(\theta,\phi)\Big),
\end{equation}
with $R(\theta,\phi)$ the radius of the nucleus, $Y_{\lambda\mu}(\theta,\phi)$ the spherical harmonics and $\alpha_{\lambda\mu}$ the collective variables of order $\lambda$ and projection $\mu$.   Up to quadrupole deformations, the surface (\ref{ellipsoid:equation}) is restricted to ellipsoidal deformations, determined by the variables $\alpha_{2\mu}$, as long as they are chosen sufficiently small ($\alpha_{2\mu}\ll1$).  The monopole variable $\alpha_{00}$ is fixed by imposing volume conservation and the dipole variables $\alpha_{1\mu}$ describe a translation of the centre of mass and are therefore neglected \cite{eisenberg:87}.  As the collective quadrupole variables\footnote{which we will abbreviate to $\alpha_\mu$ from now on.} $\alpha_{2\mu}$ are considered to be small, the potential energy term in the Hamiltonian can be described as an angular momentum scalar Taylor expansion in the collective variables, as proposed by the Frankfurt group \cite{eisenberg:87,gneuss:71,hess:80a,hess:80b,troltenier:91}.  The Hamiltonian is given by
\begin{equation}\label{hamiltonian:hamiltonian}
\hat{H}=\hat{T} + V(\alpha),
\end{equation}
with the potential energy term $V(\alpha)$,
\begin{eqnarray}
V(\alpha)&=C_2\alpha\cdot\alpha+C_3[\alpha\alpha]^2\cdot\alpha+C_4(\alpha\cdot\alpha)^2\nonumber\\
&\qquad +C_5([\alpha\alpha]^2\cdot\alpha)(\alpha\cdot\alpha)+C_6(\alpha\cdot\alpha)^3+D_6([\alpha\alpha]^2\cdot\alpha)^2+\dots.
\end{eqnarray}
where the dot denotes angular momentum scalar coupling $a_l\cdot b_l=(-)^l\sqrt{2l+1}[a_l b_l]^{(0)}_0$.  The kinetic energy term can be expanded in a similar way, through inclusion of higher order quadratic terms in the canonic momentum $\pi_\mu$ can be included, thus\footnote{Although the quadrupole variable $\alpha_\mu$ and $\pi_\nu$ are operators, we omit the operator sign to avoid notational overload.}
\begin{equation}
\hat{T}=\case{1}{2B_2}\pi\cdot\pi+B_3([\pi\alpha]^2\cdot\pi+\textrm{h.c.})+\dots.
\end{equation}
The canonic conjugate momenta are defined by the standard relations \cite{eisenberg:87}
\begin{equation}\label{variables:commutationrelations}
[\pi_{\mu^\prime},\alpha_\mu]=-i\hbar\delta_{\mu\mu^\prime},\qquad [\pi_{\mu^\prime},\pi_\mu]=0,\qquad [\alpha_{\mu^\prime},\alpha_\mu]=0.
\end{equation}
To establish the algebraic structure of the collective quadrupole model, it is convenient to change to a bosonic representation where the spin 2 phonon creation and annihilation operators are defined as
\begin{equation}\label{phonon:definition}
b_\mu^\dag=\case{1}{\sqrt{2}}(\sqrt{k}\alpha_\mu+\case{i}{\sqrt{k}\hbar}\pi_\mu^\ast),\quad \tilde{b}_\mu=\case{1}{\sqrt{2}}(\sqrt{k}\alpha_\mu-\case{i}{\sqrt{k}\hbar}\pi_\mu^\ast),
\end{equation}
with $[b_\mu,b^\dag_\nu]=\delta_{\mu\nu}$, $\tilde{b}_\mu=(-)^\mu b_{-\mu}$ and $k$ a free parameter.  On the one hand, the following 10 operators
\begin{eqnarray}
&L_M=\sqrt{10}[b^\dag\tilde{b}]^{(1)}_M\equiv\case{-i\sqrt{10}}{\hbar}[\alpha\pi^\ast]^{(1)}_M,\\
&O_M=\sqrt{10}[b^\dag\tilde{b}]^{(3)}_M\equiv\case{-i\sqrt{10}}{\hbar}[\alpha\pi^\ast]^{(3)}_M,
\end{eqnarray}
close under the commutation relations of an $O(5)$ algebra.
\begin{eqnarray}
&[L_m,L_{m^\prime}]=-\sqrt{2}\langle 1m1m^\prime|1m+m^\prime\rangle L_{m+m^\prime},\\{}
&[L_m,O_{m^\prime}]=-2\sqrt{3}\langle 1m3m^\prime|3m+m^\prime\rangle O_{m+m^\prime},\\{}
&[O_m,O_{m^\prime}]=-2\sqrt{7}\langle 3m3m^\prime|1m+m^\prime\rangle L_{m+m^\prime}\nonumber\\{}
&\qquad\qquad\qquad+\sqrt{6}\langle 3m3m^\prime|3m+m^\prime\rangle O_{m+m^\prime}.
\end{eqnarray}
On the other hand, if we define the following operators \cite{ui:68,arima:76}
\begin{equation}\label{su11:realisation}
B_+=\case{1}{2}b^\dag\cdot b^\dag,\quad B_-=\case{1}{2}\tilde{b}\cdot \tilde{b},\quad B_0=\case{1}{4}(b^\dag\cdot\tilde{b}+\tilde{b}\cdot b^\dag),
\end{equation}
they immediately give rise to an $SU(1,1)$ algebra with the commutation relations
\begin{equation}
[B_0,B_\pm]=\pm B_\pm,\qquad [B_-,B_+]=2B_0.
\end{equation}
At this point it is straightforward to see why collective quadrupole Hamiltonians (\ref{hamiltonian:hamiltonian}) can be handled within an $SU(1,1)\times O(5)$ algebraic structure.  On the one hand, the quadratic terms in $\alpha_\mu$ and $\pi_\mu$ appearing in the Hamiltonian can be written as a function of the $SU(1,1)$ generators, defined by eqs. (\ref{su11:realisation}).  Moreover, in the limiting case of a 5D harmonic oscillator system, the Hamiltonian is identical to the generator $B_0$ which is trivially diagonal in the $SU(1,1)$ basis (see section \ref{section:basis}, eqs. (\ref{su11:basis:casimir}) and (\ref{su11:basis:cartan})).  On the other, the cubic terms in the Hamiltonian (\ref{hamiltonian:hamiltonian}) cannot be recognised as generators of any known algebra.  However, because $\alpha_\mu$ and $\pi_\mu$ have good tensorial properties with respect to $O(5)$, it provides an interesting scheme to calculate these matrix elements, as will be shown in sections \ref{section:basis} and \ref{section:matrixelements}. 
\section{Establishing the basis.}\label{section:basis}
%================================
%%
To perform numerical calculations, a basis for the $SU(1,1)$ as well as the $O(5)$ basis is needed.  For the $SU(1,1)$ part, a suitable basis $|n,\lambda\rangle$ is known which diagonalises the Casimir operator $\mathcal{C}_2[SU(1,1)]$ and the Cartan operator $B_0$ \cite{rowe:96}
\begin{eqnarray}
\mathcal{C}_2[SU(1,1)]|n,\lambda\rangle=\case{1}{4}\lambda(\lambda-2)|\lambda,n\rangle,\qquad & (\lambda\in\mathbb{R}^+),\label{su11:basis:casimir}\\
B_0|n,\lambda\rangle=\case{1}{2}(\lambda+2n)|\lambda,n\rangle,\qquad & (n\in\mathbb{N})\label{su11:basis:cartan}.
\end{eqnarray}
Although in principle $\lambda$ can take any positive real value, it will be restricted to a discrete number of values, due to the explicit physical realisation of the algebra (\ref{su11:realisation}) as will be demonstrated further on in eq. (\ref{casimirrelation:operators}).  The operators $B_+$ and $B_-$ respectively act as $n$- raising and lowering operators in this basis
\begin{eqnarray}
B_+|\lambda,n\rangle=\sqrt{(\lambda+n)(n+1)}|\lambda,n+1\rangle,\\
B_-|\lambda,n\rangle=\sqrt{(\lambda+n-1)n}|\lambda,n-1\rangle.
\end{eqnarray}
Within the present work, the $O(5)$ structure will be treated in the Cartan-Weyl basis, which is defined by the following rotation of the $L_M$ and $O_M$ generators \cite{corrigan:76,debaerdemacker:07a}
\begin{equation}\label{generators:rotationtocartan}
\eqalign{
X_+=-\case{1}{5}(\sqrt{2}L_{+1}+\sqrt{3}O_{+1}),\qquad & Y_+=-\case{1}{\sqrt{5}}O_{+3},\\
X_-=\case{1}{5}(\sqrt{2}L_{-1}+\sqrt{3}O_{-1}), & Y_-=\case{1}{\sqrt{5}}O_{-3},\\
X_0=\case{1}{10}(L_0+3O_0), & Y_0=\case{1}{10}(3L_0-O_0),\\
T_{\frac{1}{2}\frac{1}{2}}=\case{1}{\sqrt{10}}O_{+2}, & T_{-\frac{1}{2}\frac{1}{2}}=-\case{1}{\sqrt{50}}(\sqrt{3}L_{+1}-\sqrt{2}O_{+1}),\\
T_{-\frac{1}{2}-\frac{1}{2}}=-\case{1}{\sqrt{10}}O_{-2}, &  T_{\frac{1}{2}-\frac{1}{2}}=\case{1}{\sqrt{50}}(\sqrt{3}L_{-1}-\sqrt{2}O_{-1}).}
\end{equation}
From these definitions, it is clear that $O(5)$ can be reduced to $O(5)\subset O(4)\cong SU(2)_X\times SU(2)Y$, as the two sets of $\{X_0,X_\pm\}$ and $\{Y_0,Y_\pm\}$ operators both span an $SU(2)$ algebra and commute with each other.  We can associate a basis $|vXM_XM_Y\rangle$ with this reduction scheme, where $v$ is defined by means of the Casimir operator of $O(5)$ ($\mathcal{C}_2[O(5)]$), $X$ by the Casimir operators of the two $SU(2)$ subgroups ($\mathcal{C}_2[SU(2)_X]\equiv\mathcal{C}_2[SU(2)_Y]$ for symmetric representation \cite{corrigan:76,debaerdemacker:07a}) and $\{M_X,M_Y\}$ by the Cartan subalgebra $\{X_0,Y_0\}$
\begin{eqnarray}
\mathcal{C}_2[O(5)]|vXM_XM_Y\rangle=v(v+3)|vXM_XM_Y\rangle,\\
\mathcal{C}_2[SU(2)_{X,Y}]|vXM_XM_Y\rangle=X(X+1)|vXM_XM_Y\rangle,\\
X_0|vXM_XM_Y\rangle=M_X|vXM_XM_Y\rangle,\\
Y_0|vXM_XM_Y\rangle=M_Y|vXM_XM_Y\rangle.
\end{eqnarray}
The seniority quantum number $v$ is an integer number, $X=0\dots v/2$, and the quantum numbers $\{M_X,M_Y\}$ are defined through the standard $SU(2)$ reduction rules ($M_X=-X\dots X$, $M_Y=-X\dots X$).  The action of the $SU(2)_X$ and $SU(2)_Y$ generators on the Cartan-Weyl basis are well-known from angular momentum theory \cite{rose:57} whereas the action of the $T_{\mu\nu}$ generators can be obtained by means of an intermediate state method, fully exploiting the bispinorial character of the generators within the Cartan-Weyl framework \cite{debaerdemacker:07a}.  The essential results of this calculation are given in \ref{appendix:A}.\newline
Although the generators of $O(5)$ commute with all generators of $SU(1,1)$, the algebras are still connected through the Casimir operators
\begin{equation}\label{casimirrelation:operators}
\mathcal{C}_2[SU(1,1)]=\case{1}{4}(\mathcal{C}_2[O(5)]+\case{5}{4}),
\end{equation}
with the Casimir operators defined by
\begin{equation}
\mathcal{C}_2[SU(1,1)]=B_0^2-B_0-B_+B_-,\qquad
\mathcal{C}_2[O(5)]=\case{1}{5}(L\cdot L+O\cdot O).
\end{equation}
Acting with the identity (\ref{casimirrelation:operators}) on a product basis $|n\lambda\rangle|vXM_XM_Y\rangle$, we obtain a relationship between the Casimir quantum numbers
\begin{equation}\label{casimirrelation:quantumnumbers}
\lambda(\lambda-2)=v(v+3)+\case{5}{4}\rightarrow \lambda=v+\case{5}{2},
\end{equation}
as only real positive values of $\lambda$ can define unitary representations of $SU(1,1)$ \cite{rowe:96}.  This result is generally true for any dimension $N$ \cite{rowe:05b}, where $SU(1,1)\times O(N)$ is embedded in a larger non-compact symplectic group $Sp(N,\mathbb{R})$, i.e.,
\begin{equation}
\lambda=v+\case{N}{2},
\end{equation}
with $\lambda$ and $v$ the Casimir quantum numbers of the $SU(1,1)$ and $O(N)$ groups respectively.  The non-compactness stems from the fact that the raising and lowering operators of the $Sp(N,\mathbb{R})$ algebra connect different representations of the $Sp(N,\mathbb{R})\supset U(N)\supset O(N)$ group reduction, associated with harmonic oscillations in $N$ dimensions.  This non-compactness is inherited by the $SU(1,1)$ subgroup in the $Sp(N,\mathbb{R})\supset SU(1,1)\times O(N)$ reduction.   It can also be shown that relation (\ref{casimirrelation:quantumnumbers}) is the spherical harmonic oscillator limit of a more general system with deformation-driving Davidson interactions \cite{rowe:98}.  This is due to the fact that the inclusion of the centrifugal-like term in the Davidson interaction can be incorporated within the $SU(1,1)\times O(N)$ algebraic framework of the harmonic oscillator.\newline
The main consequence of equation (\ref{casimirrelation:quantumnumbers}) is that we can merge $SU(1,1)$ and $O(5)$ into one basis, defined by the quantum numbers $|nvXM_XM_Y\rangle$, tacitly omitting the quantum number $\lambda$ which is in one-to-one correspondence with the seniority quantum number $v$ by means of relation (\ref{casimirrelation:quantumnumbers}).  It turns out that this basis is very convenient to calculate the matrix elements of collective quadrupole variables and canonic conjugate momenta since it leads towards a matrix representation of collective Hamiltonians, suitable for numerical diagonalisation. 
\section{Matrix elements of the phonon creation and annihilation operators.}\label{section:matrixelements}
%%=========================================================================
%%
In this section, we calculate the matrix elements of the phonon creation and annihilation operators (\ref{phonon:definition}) in the natural Cartan-Weyl basis of $SU(1,1)\times O(5)$.  At first, we only consider the $SU(1,1)$ contribution and incorporate the $O(5)$ Cartan-Weyl representations later.  To avoid notational overload in the first part, we initially abbreviate the full basis state $|nvXM_XM_Y\rangle$ by its $SU(1,1)$-part $|n\lambda\rangle$ (and $|n^\prime v^\prime X^\prime M_X^\prime M_Y^\prime \rangle$ by $|n^\prime \lambda^\prime \rangle$).  This is allowed as long as only the $SU(1,1)$ generators are involved.  Whenever the $O(5)$ generators are incorporated, the full notation is required.\newline
First, we derive selection rules for the matrix elements in the $SU(1,1)$ basis.  For this purpose, we write down the commutation relations
\begin{equation}
	\begin{array}{ll}
	[B_-,b_\mu^\dag]=\tilde{b}_\mu,&[B_-,\tilde{b}_\mu]=0,\\{}
	[B_0,b_\mu^\dag]=\case{1}{2}b^\dag_\mu,&[B_0,\tilde{b}_\mu]=-\case{1}{2}\tilde{b}_\mu,\\{}
	[B_+,b_\mu^\dag]=0,&[B_+,\tilde{b}_\mu]=-b^\dag_\mu.
	\end{array}
\end{equation}
The calculation of the matrix elements of the commutation relations with $B_0$ 
\begin{eqnarray}
&\langle \lambda^\prime n^\prime|[B_0,b_\mu^\dag]|\lambda n\rangle=\case{1}{2}\langle \lambda^\prime n^\prime|b_\mu^\dag|\lambda n\rangle,\\
&\langle \lambda^\prime n^\prime |[B_0,\tilde{b}_\mu]|\lambda n\rangle=-\case{1}{2}\langle \lambda^\prime n^\prime|\tilde{b}_\mu|\lambda n\rangle,
\end{eqnarray}
results into to the selection rules 
\begin{eqnarray}
&\lambda^\prime+2n^\prime-\lambda-2n-1=0\qquad\textrm{for }\langle\lambda^\prime n^\prime|b_\mu^\dag|\lambda n\rangle,\label{boson:selectionrule1a}\\
&\lambda^\prime+2n^\prime-\lambda-2n+1=0\qquad\textrm{for }\langle\lambda^\prime n^\prime|\tilde{b}_\mu|\lambda n\rangle.\label{boson:selectionrule1b}
\end{eqnarray}
More selection rules can be obtained from the other commutation relations
\begin{eqnarray}
&\langle \lambda^\prime n^\prime+1|[B_+,b_\mu^\dag]|\lambda n-1\rangle=0,\\
&\langle \lambda^\prime n^\prime+1|[B_+,\tilde{b}_\mu]|\lambda n\rangle=-\langle\lambda^\prime n^\prime+1|b_\mu^\dag|\lambda n\rangle,\\
&\langle \lambda^\prime n^\prime|[B_-,b_\mu^\dag]|\lambda n\rangle=\langle\lambda^\prime n^\prime|\tilde{b}_\mu|\lambda n\rangle,\\
&\langle \lambda^\prime n^\prime|[B_-,\tilde{b}_\mu]|\lambda n+1\rangle=0.
\end{eqnarray}
The above relations give rise to the following set of four equations
\begin{eqnarray}
&\sqrt{(\lambda^\prime+n^\prime)(n^\prime+1)}\langle\lambda^\prime n^\prime|b_\mu^\dag|\lambda n-1\rangle-\sqrt{(\lambda+n-1)n}\langle\lambda^\prime n^\prime+1|b_\mu^\dag|\lambda n\rangle\nonumber\\
&\qquad=0,\\
&\sqrt{(\lambda^\prime+n^\prime)(n^\prime+1)}\langle\lambda^\prime n^\prime|\tilde{b}_\mu|\lambda n\rangle-\sqrt{(\lambda+n)(n+1)}\langle\lambda^\prime n^\prime+1|\tilde{b}_\mu|\lambda n+1\rangle\nonumber\\
&\qquad=-\langle\lambda^\prime n^\prime+1|b_\mu^\dag|\lambda n\rangle,\\
&\sqrt{(\lambda^\prime+n^\prime)(n^\prime+1)}\langle\lambda^\prime n^\prime+1|b_\mu^\dag|\lambda n\rangle-\sqrt{(\lambda+n-1)n}\langle\lambda^\prime n^\prime|b_\mu^\dag|\lambda n-1\rangle\nonumber\\
&\qquad=\langle\lambda^\prime n^\prime|\tilde{b}_\mu|\lambda n\rangle,\\
&\sqrt{(\lambda^\prime+n^\prime)(n^\prime+1)}\langle\lambda^\prime n^\prime+1|\tilde{b}_\mu|\lambda n+1\rangle-\sqrt{(\lambda+n)(n+1)}\langle\lambda^\prime n^\prime|\tilde{b}_\mu|\lambda n\rangle\nonumber\\
&\qquad=0,
\end{eqnarray}
which forms a homogeneous set of four equations in four variables (the matrix elements).  Therefore, these matrix elements are identically zero unless the determinant of the matrix vanishes.  Solving the singularity equation, keeping in mind that the four matrix elements are chosen in such a way that the selection rule $\lambda^\prime+2n^\prime-\lambda-2n+1=0$ holds (see eqs. \ref{boson:selectionrule1a} and \ref{boson:selectionrule1b}) for all of them, we obtain the general selection rules
\begin{eqnarray}
&\{\lambda^\prime=\lambda-1,n^\prime=n\},\label{boson:selectionrule2a}\\
&\{\lambda^\prime=\lambda+1,n^\prime=n-1\}\label{boson:selectionrule2b}.
\end{eqnarray}
Thus the non-vanishing matrix elements are
\begin{equation}
\begin{array}{ll}
\langle\lambda+1,n|b_\mu^\dag|\lambda n\rangle, & \langle\lambda+1,n-1|\tilde{b}_\mu|\lambda n\rangle, \\
\langle\lambda-1,n+1|b_\mu^\dag|\lambda n\rangle, & \langle\lambda-1,n|\tilde{b}_\mu|\lambda n\rangle.
\end{array}
\end{equation}
Although we already knew that $\alpha_\mu$ and $\pi^\ast_\mu$ (and subsequently $b_\mu^\dag$ and $\tilde{b}_\mu$) are $v=1$ $O(5)$ tensors, we stress that this was not explicitly taken into account in the above calculation, but emerged naturally from the selection criteria.  However, these selection rules also contain a physical interpretation.  It is clear from the definition of the raising operator $B_+$ that the quantum number $n$ denotes the number of phonon pairs coupled to angular momentum zero.  Since $B_0$ counts the total number of phonons, $\lambda$ can be associated with the number of pairs not coupled to zero, i.e. the seniority $v$.  Therefore, creating or annihilating a single phonon, by means of $b^\dag_\mu$ and $\tilde{b}_\mu$ respectively, can result in the creation or annihilation of a pair, provided that the total number of phonons ($2n+v$) is increased or decreased by one.  To conclude, the allowed solutions of the homogeneous set of equations are summarised.  Some solutions relate the matrix elements with different pair number $n$
\begin{eqnarray}
&\sqrt{\lambda+n+1}\langle\lambda+1,n|b^\dag_\mu|\lambda n\rangle=\sqrt{\lambda+n}\langle\lambda+1,n+1|b^\dag_\mu|\lambda n+1\rangle,\label{boson:rule3a}\\
&\sqrt{n+1}\langle\lambda-1,n|b^\dag_\mu|\lambda, n-1\rangle=\sqrt{n}\langle\lambda-1,n+1|b^\dag_\mu|\lambda n\rangle,\label{boson:rule3b}
\end{eqnarray}
and
\begin{eqnarray}
&\sqrt{n}\langle\lambda+1,n|\tilde{b}_\mu|\lambda n+1\rangle=\sqrt{n+1}\langle\lambda+1,n-1|\tilde{b}_\mu|\lambda n\rangle,\label{boson:rule4a}\\
&\sqrt{\lambda+n-1}\langle\lambda-1,n+1|\tilde{b}_\mu|\lambda, n+1\rangle=\sqrt{\lambda+n}\langle\lambda-1,n|\tilde{b}_\mu|\lambda n\rangle,\label{boson:rule4b}
\end{eqnarray}
while other relate the matrix elements of creation and annihilation operators
\begin{eqnarray}
&\sqrt{n}\langle\lambda+1,n|b_\mu^\dag|\lambda n\rangle=\sqrt{\lambda+n}\langle\lambda+1,n-1|\tilde{b}_\mu|\lambda n\rangle,\label{boson:rule5a}\\
&\sqrt{\lambda+n-1}\langle\lambda-1,n+1|b_\mu^\dag|\lambda n\rangle=\sqrt{n+1}\langle\lambda-1,n|\tilde{b}_\mu|\lambda n\rangle.\label{boson:rule5b}
\end{eqnarray}
From this point onwards, we include the $O(5)$ basis explicitly.  As the creation and annihilation operator matrix elements can be related by means of equation (\ref{boson:rule5a}) and (\ref{boson:rule5b}), we only consider $b^\dag$.  Relying on Racah's theorem \cite{racah:42}, we know that the collective variables $\alpha_\mu$ can be classified as a single biscalar part $\{00\}$ and the four components of a bispinor with respect to the $SU(2)_X\times SU(2)_Y$ subalgebra of $O(5)$ \cite{corrigan:76,debaerdemacker:07a}, i.e.
\begin{eqnarray}
&[X_{0},\alpha^{\lambda\lambda}_{\mu\nu}]=\mu \alpha^{\lambda\lambda}_{\mu\nu},\\
&[X_{\pm},\alpha^{\lambda\lambda}_{\mu\nu}]=\sqrt{(\lambda\mp\mu)(\lambda\pm\mu+1)}\alpha^{\lambda\lambda}_{\mu\pm1\nu},\\
&[Y_{0},\alpha^{\lambda\lambda}_{\mu\nu}]=\nu \alpha^{\lambda\lambda}_{\mu\nu},\\
&[Y_{\pm},\alpha^{\lambda\lambda}_{\mu\nu}]=\sqrt{(\lambda\mp\nu)(\lambda\pm\nu+1)}\alpha^{\lambda\lambda}_{\mu\nu\pm1},
\end{eqnarray}
where the 5 collective variables have been relabelled as follows
\begin{eqnarray}
&\Big\{\alpha_2=\alpha_{\frac{1}{2}\frac{1}{2}}^{\frac{1}{2}\frac{1}{2}}, \alpha_1=\alpha_{-\frac{1}{2}\frac{1}{2}}^{\frac{1}{2}\frac{1}{2}}, \alpha_{-1}=\alpha_{\frac{1}{2}-\frac{1}{2}}^{\frac{1}{2}\frac{1}{2}}, \alpha_{-2}=\alpha_{-\frac{1}{2}-\frac{1}{2}}^{\frac{1}{2}\frac{1}{2}}\Big\},\nonumber\\
&\Big\{\alpha_0=\alpha_{00}^{00}\Big\}.\label{variables:bitensorcharacter}
\end{eqnarray}
Remarkably, the same classification can be carried out for the canonic conjugate momenta $\pi^\ast_\mu$, implying that the phonon creation and annihilation operators can also be classified according to the bitensorial character with respect to $SU(2)_X\times SU(2)_Y$.  Therefore, we can relabel the operators as
\begin{eqnarray}
&\Big\{{b^\dag}_2={b^\dag}_{\frac{1}{2}\frac{1}{2}}^{\frac{1}{2}\frac{1}{2}}, {b^\dag}_1={b^\dag}_{-\frac{1}{2}\frac{1}{2}}^{\frac{1}{2}\frac{1}{2}}, {b^\dag}_{-1}={b^\dag}_{\frac{1}{2}-\frac{1}{2}}^{\frac{1}{2}\frac{1}{2}}, {b^\dag}_{-2}={b^\dag}_{-\frac{1}{2}-\frac{1}{2}}^{\frac{1}{2}\frac{1}{2}}\Big\},\\
&\Big\{{b^\dag}_0={b^\dag}_{00}^{00}\Big\}.
\end{eqnarray}
Following the same procedure as described in \cite{debaerdemacker:07a}, we can separate out the projection quantum numbers $M_X$ and $M_Y$, making use of the double reduced matrix elements\footnote{We formally use the single reduced matrix notation in order to express the double reduced matrix elements, as any confusion between normal and double reduced matrix elements is excluded within this work.}, defined by the Wigner-Eckart theorem 
\begin{eqnarray}
&\langle n v X M_XM_Y|{b^\dag}^{\lambda\lambda}_{\mu\nu}|n^\prime v^\prime X^\prime M_X^\prime M_Y^\prime\rangle\\
&=(-)^{\phi}\wigner{X}{\lambda}{X^\prime}{-M_X}{\mu}{M_X^\prime}\wigner{X}{\lambda}{X^\prime}{-M_Y}{\nu}{M_Y^\prime}\langle nvX||{b^\dag}^\lambda||n^\prime v^\prime X^\prime\rangle,\nonumber
\end{eqnarray} 
with $\phi=2X-M_X-M_Y$.\newline
We can now calculate the double reduced matrix elements of the phonon creation operators.  For this purpose, we will follow the same algorithm as described in \cite{debaerdemacker:07a}: calculate the expectation value of the necessary commutation relations in the $SU(1,1)\times O(5)$ basis and insert a complete set of intermediate basis states.  This results in an overdetermined set of equations which can be solved algebraically with respect to the needed matrix elements.\newline
We start from the commutation relations of the phonon creation operators with the $T_{\mu\nu}$ generators
\begin{eqnarray}
&[T_{\mu\nu},{b^\dag}_{\mu^\prime\nu^\prime}^{\frac{1}{2}\frac{1}{2}}]=\case{(-)^{(\mu+\nu)}}{\sqrt{2}}\delta_{-\mu\mu^\prime}\delta_{-\nu\nu^\prime}{b^\dag}_{00}^{00},\\{}
&[T_{\mu\nu},{b^\dag}_{00}^{00}]=\case{1}{\sqrt{2}}{b^\dag}_{\mu\nu}^{\frac{1}{2}\frac{1}{2}}.
\end{eqnarray}
As the $T_{\mu\nu}$ generators do not affect the $SU(1,1)$ quantum numbers, we can repeat the procedure outlined in \cite{debaerdemacker:07a}, from step (54) until (64), to obtain expressions that relate the double reduced matrix elements of ${b^\dag}^{\frac{1}{2}}$ with those of ${b^\dag}^{0}$.  Moreover, we can recover the seniority selection rules $v^\prime=v\pm1$, compatible with the selection rules for $\lambda$ obtained in the previous derivation (see eqs. \ref{boson:selectionrule2a} and \ref{boson:selectionrule2b}).  If we explicitly take these selection rules into account, we obtain for $v^\prime=v+1$
\begin{eqnarray}
&\langle nv,X+\case{1}{2}||{b^\dag}^{\frac{1}{2}}||n-1,v+1,X\rangle\nonumber\\
&\qquad=-\case{1}{\sqrt{2}}\sqrt{\case{2X+2}{2X+1}}\sqrt{\case{v-2X}{v+2X+3}}\langle nvX||{b^\dag}^0||n-1,v+1,X\rangle,\label{boson:rule6a}\\
&\langle nv,X||{b^\dag}^{\frac{1}{2}}||n-1,v+1,X+\case{1}{2}\rangle\nonumber\\
&\qquad=\case{1}{\sqrt{2}}\sqrt{\case{2X+2}{2X+1}}\sqrt{\case{v+2X+4}{v-2X+1}}\langle nvX||{b^\dag}^0||n-1,v+1,X\rangle,\label{boson:rule6b}
\end{eqnarray}
and for $v^\prime=v-1$
\begin{eqnarray}
&\langle nv,X+\case{1}{2}||{b^\dag}^{\frac{1}{2}}||n,v-1,X\rangle\nonumber\\
&\qquad=\case{1}{\sqrt{2}}\sqrt{\case{2X+2}{2X+1}}\sqrt{\case{v+2X+3}{v-2X}}\langle nvX||{b^\dag}^0||n,v-1,X\rangle,\label{boson:rule7a}\\
&\langle nv,X||{b^\dag}^{\frac{1}{2}}||n,v-1,X+\case{1}{2}\rangle\nonumber\\
&\qquad=-\case{1}{\sqrt{2}}\sqrt{\case{2X+2}{2X+1}}\sqrt{\case{v-2X-1}{v+2X+2}}\langle nvX||{b^\dag}^0||n,v-1,X\rangle.\label{boson:rule7b}
\end{eqnarray}
Hence, we only need to determine two double reduced $\{00\}$ matrix elements.  To do so, we have two additional expressions at hand: the commutation relation $[\tilde{b}_0,b^\dag_0]=1$ and the $SU(1,1)$ generator $B_+=\frac{1}{2}b^\dag\cdot b^\dag$.  We construct the following matrix elements
\begin{eqnarray}
&\langle nvXM_XM_Y|[\tilde{b}_0,b^\dag_0]|nvXM_XM_Y\rangle=1,\\
&\langle n+1,vXM_XM_Y|b^\dag\cdot b^\dag|nvXM_XM_Y\rangle=2\sqrt{(\lambda+n)(n+1)}.
\end{eqnarray}
Going over to double reduced matrix elements and applying the intermediate state method while making use of the relations (\ref{boson:rule3a} to \ref{boson:rule5b}) and (\ref{boson:rule6a} to \ref{boson:rule7b}), we obtain the closed results
\begin{eqnarray}
\langle n+1 vX||&{b^\dag}^0||n,v+1,X\rangle\langle n,v+1,X||{b^\dag}^0||nvX\rangle\nonumber\\
&=\case{(v+2X+3)(v-2X+1)}{(2v+5)(2v+3)}2(2X+1)^2\sqrt{(\lambda+n)(n+1)},\\
\langle n+1 vX||&{b^\dag}^0||n+1,v-1,X\rangle\langle n+1,v-1,X||{b^\dag}^0||nvX\rangle\nonumber\\
&=\case{(v+2X+2)(v-2X)}{(2v+1)(2v+3)}2(2X+1)^2\sqrt{(\lambda+n)(n+1)}.
\end{eqnarray}
Since $(b^\dag_0)^\dag=b_0\equiv\tilde{b}_0$, we obtain the boson creation double reduced matrix elements
\begin{eqnarray}
&\langle n,v+1,X||{b^\dag}^0||nvX\rangle=\sqrt{\case{(v+2X+3)(v-2X+1)}{(2v+5)(2v+3)}}\sqrt{2}(2X+1)\sqrt{\lambda+n},\label{boson:closedexpression1}\\
&\langle n+1,v-1,X||{b^\dag}^0||nvX\rangle=\sqrt{\case{(v+2X+2)(v-2X)}{(2v+1)(2v+3)}}\sqrt{2}(2X+1)\sqrt{n+1},\label{boson:closedexpression2}
\end{eqnarray}
and likewise for the boson annihilation double reduced matrix elements
\begin{eqnarray}
&\langle n-1,v+1,X||\tilde{b}^0||nvX\rangle=\sqrt{\case{(v+2X+3)(v-2X+1)}{(2v+5)(2v+3)}}\sqrt{2}(2X+1)\sqrt{n},\label{boson:closedexpression3}\\
&\langle n,v-1,X||\tilde{b}^0||nvX\rangle=\sqrt{\case{(v+2X+2)(v-2X)}{(2v+1)(2v+3)}}\sqrt{2}(2X+1)\sqrt{\lambda+n-1}.\label{boson:closedexpression4}
\end{eqnarray}
Finally, we obtain closed expressions for all phonon creation and annihilation operators in the $SU(1,1)\times O(5)$ Cartan-Weyl basis.  Since the collective quadrupole variables $\alpha_\mu$, as well as the canonic conjugate momenta can be expressed as a function of the phonon creation and annihilation operators (\ref{phonon:definition}), the full collective Hamiltonian (\ref{hamiltonian:hamiltonian}) can be expressed as a matrix representation in the natural basis.  However, the Cartan-Weyl reduction is not naturally compatible with the physical angular momentum quantum number $L$, which emerges from experimental energy spectra of atomic nuclei.  Therefore, one needs to rotate the Cartan-Weyl basis to the physical basis by diagonalising the operator $L\cdot L$.  At this point, it is unclear whether this rotation can be carried out analytically.  Hence, this rotation is performed numerically in actual calculations \cite{debaerdemacker:07a}.\newline
\section{Conclusions}
Collective modes of motion are of utmost importance in the low-energy spectra of atomic nuclei.  Therefore, a good scheme to diagonalise the Bohr-Hamiltonian (\ref{hamiltonian:hamiltonian}) is required.  In the present work, we have shown that it is convenient to construct a matrix representation of the Hamiltonian in the Cartan-Weyl basis of $SU(1,1)\times O(5)$, since closed expressions of all matrix elements of the basis quadrupole variables $\alpha_\mu$ and canonic conjugate momenta $\pi_\mu$ can be obtained algebraically.  As a result, we can now study general collective modes of motion and test their applicability to the description of experimental nuclear structure data, in particular with respect to the recent developments in rare-isotope facilities.  Some first exploratory results have been obtained \cite{debaerdemacker:07b} within the framework of quantum shape phase transitions \cite{iachello:00,iachello:01,iachello:03}, and will be discussed extensively in a forthcoming publication. 
\section*{Acknowledgements}
%==========================
The authors like to thank, P. Van Isacker, J. Van der Jeugt and R. Campoamor-Herzberg for interesting discussions and suggestions.  Financial support from the University of Ghent, the ''FWO-Vlaanderen'' and the Interuniversity Attraction Pool (IUAP) under project P6/23 that made this research possible is acknowledged.   We would also like to thank the [Department of Energy's] Institute for Nuclear Theory at the University of Washington for its hospitality and the Department of Energy for partial support during the completion of this work.
\appendix
%=======
%
%
\section{Action of the $O(5)$ generators on the Cartan-Weyl basis}\label{appendix:A}
%==================================================
%
The action of the $SU(2)_X$ and $SU(2)_Y$ generators are trivial, thanks to the well-known angular momentum theory \cite{rose:57}
\begin{eqnarray}
X_{\pm}|vXM_XM_Y\rangle&=\sqrt{(X\mp M_X)(X\pm M_X+1)}|vXM_X\pm1,M_Y\rangle,\\
X_{0}|vXM_XM_Y\rangle&=M_X|vXM_XM_Y\rangle,\\
Y_{\pm}|vXM_XM_Y\rangle&=\sqrt{(X\mp M_Y)(X\pm M_Y+1)}|vXM_X,M_Y\pm1\rangle,\\
Y_{0}|vXM_XM_Y\rangle&=M_Y|vXM_XM_Y\rangle.
\end{eqnarray}
The $T_{\mu\nu}$ are less trivial, but can be obtained by means of an intermediate state method \cite{debaerdemacker:07a}
\begin{eqnarray}
\eqalign{\fl T_{\frac{1}{2}\frac{1}{2}}|vXM_XM_Y\rangle=\\
\fl\quad\frac{\sqrt{(X+M_X+1)(X+M_Y+1)(v-2X)(v+2X+3)}}{2\sqrt{(2X+1)(2X+2)}}|vX+\case{1}{2},M_X+\case{1}{2},M_Y+\case{1}{2}\rangle\label{Tgenerator:plusplus}\\
\fl\quad-\frac{\sqrt{(X-M_X)(X-M_Y)(v-2X+1)(v+2X+2)}}{2\sqrt{(2X)(2X+1)}}|vX-\case{1}{2},M_X+\case{1}{2},M_Y+\case{1}{2}\rangle,}\\
\eqalign{\fl T_{\frac{1}{2}-\frac{1}{2}}|vXM_XM_Y\rangle=\\
\fl\quad=\frac{\sqrt{(X+M_X+1)(X-M_Y+1)(v-2X)(v+2X+3)}}{2\sqrt{(2X+1)(2X+2)}}|vX+\case{1}{2},M_X+\case{1}{2},M_Y-\case{1}{2}\rangle\\
\fl\quad+\frac{\sqrt{(X-M_X)(X+M_Y)(v-2X+1)(v+2X+2)}}{2\sqrt{(2X)(2X+1)}}|vX-\case{1}{2},M_X+\case{1}{2},M_Y-\case{1}{2}\rangle,}\\
\eqalign{\fl T_{-\frac{1}{2}\frac{1}{2}}|vXM_XM_Y\rangle\\
\fl\quad=\frac{\sqrt{(X-M_X+1)(X+M_Y+1)(v-2X)(v+2X+3)}}{2\sqrt{(2X+1)(2X+2)}}|vX+\case{1}{2},M_X-\case{1}{2},M_Y+\case{1}{2}\rangle\\
\fl\quad+\frac{\sqrt{(X+M_X)(X-M_Y)(v-2X+1)(v+2X+2)}}{2\sqrt{(2X)(2X+1)}}|vX-\case{1}{2},M_X-\case{1}{2},M_Y+\case{1}{2}\rangle,}\\
\eqalign{\fl T_{-\frac{1}{2}-\frac{1}{2}}|vXM_XM_Y\rangle\\
\fl\quad=\frac{\sqrt{(X-M_X+1)(X-M_Y+1)(v-2X)(v+2X+3)}}{2\sqrt{(2X+1)(2X+2)}}|vX+\case{1}{2},M_X-\case{1}{2},M_Y-\case{1}{2}\rangle\\
\fl\quad-\frac{\sqrt{(X+M_X)(X+M_Y)(v-2X+1)(v+2X+2)}}{2\sqrt{(2X)(2X+1)}}|vX-\case{1}{2},M_X-\case{1}{2},M_Y-\case{1}{2}\rangle.}
\end{eqnarray}
\section*{References}
%================
%
%
\bibliography{debaerdemacker_quadrupolecollectivemodel}
\bibliographystyle{iopart-num}
\end{document}